\documentclass[twocolumn,preprintnumbers,amsmath,amssymb]{revtex4}
\usepackage{amssymb,amsmath,graphicx,rotating}
\begin{document}
\title{ Finding irreducible representation of symmetry operators linearly}
\author{ Young-Chung Hsue }
\affiliation{Department of Physics, National Cheng Kung University, Tainan 70101, Taiwan}
\affiliation{National Center for Theoretical Sciences, National Tsing Hua University, Hsinchu 30043, Taiwan}
\begin{abstract}
The main purpose of this paper is providing a simple method to generate the matrices of irreducible representations because it is useful to reduce the computational time of solving the eigenvalue problems. The only information we need to provide for this method is the group-multiplication table, and the proof of validity of this method is also shown in this paper.
\end{abstract}
\maketitle

\section{Introduction}
When we study the quantum mechanics~\cite{Quantum_mechanism_chinese,Quantum_mechanism_molecular}, we usually solve schr\"odinger equation or Kohn-Sham equation in many body system and we solved them through the equation $\hat H\psi=\varepsilon\psi$ where $\hat H$, $\varepsilon$ and $\psi$ are Hamiltonian, eigenvalue and eigenfunction, respectively.
In this equation, $\psi$ is expanded by an independent bases usually and $\hat H$ will be a $N\times N$ matrix.
For large $N$ case, computing the eigenvalues and eigenstates of $\hat H$ will be time consuming. 
One of the solution to save time is the use of symmetry.
The symmetry can not only be applied on the charge density and potential but also on the eigenfunctions.
However, if we want to apply the symmetry on the eigenfunctions, we need to have irreducible representation first~\cite{reduce_matrix_APW_Rudge}.
Although people already prove that one character table is related to one kind of irreducible representation, we still need to find a way to find the irreducible representation from the character table.
Besides, the equations people used to find out the character table is the
multiplication relation of $\hat \chi(\hat{R}_i)$ which is the trace of representation of $\hat{R}_i$~\cite{Group_theory}.
Unfortunately, it is not a set of linear equations and will be not so easy to find the solution numerically.
Hence, I tried to find a method which is linear and simple, and show it in the following.
Although you need to compute the eigenvalues and eigenstates of a $N\times N$ matrix where $N$ is the same as the number of symmetry operators, the irreducible representations are derived when you get the eigenstates.
The method to derive the irreducible representation is shown in Sec.~\ref{sec:irr_get_IR}.
I also show the proof of the validity of this method in Sec.~\ref{sec:irr_prove_get_IR}.
%
%
Finally, I will show a brief conclusion in Sec.~\ref{sec:conclusion}.


\section{ One simple way to derive the irreducible representations 
\label{sec:irr_get_IR}}
The irreducible representation is useful when we try to find a system's 
eigenfunctions. We can use it to predict the degeneracy and to
reduce the calculation time. Following is a simple way to derive the 
irreducible representation and character table.

\subsubsection{group-multiplication table}
First of all, we need the symmetry operators.
%
%
Once we get the symmetry operators, we can get the group-multiplication table.

For example, the square lattice case, the symmetry operators for vectors are
\begin{align*}
\hat R_1=\left(\begin{array}{rr} 1&0\\0&1 \end{array}\right),
\hat R_2=\left(\begin{array}{rr} 0&1\\-1&0 \end{array}\right),
\hat R_3=\left(\begin{array}{rr} 0&-1\\1&0 \end{array}\right),\\
\hat R_4=\left(\begin{array}{rr} -1&0\\0&-1 \end{array}\right),
\hat R_5=\left(\begin{array}{rr} 1&0\\0&-1 \end{array}\right),
\hat R_6=\left(\begin{array}{rr} 0&-1\\-1&0 \end{array}\right),\\
\hat R_7=\left(\begin{array}{rr} 0&1\\1&0 \end{array}\right),
\hat R_8=\left(\begin{array}{rr} -1&0\\0&1 \end{array}\right).
\end{align*}
%
%
Then we can get the group-multiplication table
\begin{align*}
\begin{array}{cc}
& {\rm Right} \\
\begin{rotate}{90} {\rm left} \end{rotate} &
    \begin{array}{c|cccccccc}
C_{4v}  & \hat R_1 & \hat R_2 & \hat R_3 & \hat R_4 & \hat R_5 & \hat R_6 & \hat R_7 & \hat R_8 \\
\hline\\
\hat R_1  &1&2&3&4&5&6&7&8\\
\hat R_2  &2&4&1&3&6&8&5&7\\
\hat R_3  &3&1&4&2&7&5&8&6\\
\hat R_4  &4&3&2&1&8&7&6&5\\
\hat R_5  &5&7&6&8&1&3&2&4\\
\hat R_6  &6&5&8&7&2&1&4&3\\
\hat R_7  &7&8&5&6&3&4&1&2\\
\hat R_8  &8&6&7&5&4&2&3&1
    \end{array} 
\end{array}
\end{align*}
where the members denoted by $1\to 8$ means $\hat R_1\to \hat R_8$ and they equal $\hat R_{\rm left}\hat R_{\rm right}$, respectively.
%
%

\subsubsection{$\hat{H}$ \label{sec:irr_fake_Hami}}
After we get the group-multiplication table, we can use it to generate a
Hamiltonian $\hat H$ which obeys this symmetry for finding irreducible representation.
Here I choose the plane wave as basis and define $H_{\bf G,G'}=\int e^{i({\bf G'-G})\cdot{\bf r}}H({\bf r})$.
Because of symmetry, $\hat H({\bf r})=\hat H(\hat{R}{\bf r})$, we will get
\begin{align}
H_{\hat{R}{\bf G},\hat{R}{\bf G'}}=H_{\bf G,G'}, \label{eq:H_R&H}
\end{align}
so that we can classify $H_{\bf G,G'}$ and just need to provide one value for each class.
The value I provided for each class is $H_{\bf G,G'}=1/t$ (depends on you), where $t$ means $t_{\rm th}$ class.
The value I chosen is for breaking the unnecessary degeneracy and then each kind of eigenvalue related to one kind of irreducible representation.

What is the ${\bf G}$? In fact, for deriving the irreducible representation
we don't need to know what it is, we just need the group-multiplication table.
We can define $\hat{R}_i{\bf G}_0={\bf G}_i$, so that
$H_{{\bf G}_i,{\bf G}_j}=H_{\hat{R}\hat{R}_i{\bf G}_0,\hat{R}\hat{R}_j{\bf G}_0}$
for all $\hat{R}$.
Hence, the $H_{\bf G,G'}$ for square lattice generated by this way is 
\begin{align*}
H_{\bf G,G'}=
\left(\begin{array}{cccccccc}
     1  &   1/{2}  &   1/{2}  &   1/{3}  &   1/{4}  &   1/{5}  &   1/{6}  &   1/{7} \\
     1/{2}  &   1  &   1/{3}  &   1/{2}  &   1/{6}  &   1/{4}  &   1/{7}  &   1/{5} \\
     1/{2}  &   1/{3}  &   1  &   1/{2}  &   1/{5}  &   1/{7}  &   1/{4}  &   1/{6} \\
     1/{3}  &   1/{2}  &   1/{2}  &   1  &   1/{7}  &   1/{6}  &   1/{5}  &   1/{4} \\
     1/{4}  &   1/{6}  &   1/{5}  &   1/{7}  &   1  &   1/{2}  &   1/{2}  &   1/{3} \\
     1/{5}  &   1/{4}  &   1/{7}  &   1/{6}  &   1/{2}  &   1  &   1/{3}  &   1/{2} \\
     1/{6}  &   1/{7}  &   1/{4}  &   1/{5}  &   1/{2}  &   1/{3}  &   1  &   1/{2} \\
     1/{7}  &   1/{5}  &   1/{6}  &   1/{4}  &   1/{3}  &   1/{2}  &   1/{2}  &   1
\end{array}\right),
\end{align*}
where I force $\hat H$ to be a hermitian operator, which means $H_{\bf G,G'}=H^*_{\bf G',G}$
and $\hat H({\bf r})\in {\mathbb R}$.
For example, you can find $H_{1,2}=H_{1,3}$ in this case;
the reason is $\hat{R}_3\hat{R}_2=\hat{R}_1$ which is shown in group-multiplication table and $H_{\bf G,G'}=H_{\hat{R}{\bf G},\hat{R}{\bf G'}}$,
we get $H_{1,2}=H_{3,1}$; besides, because of the requirement of hermitian,
$H_{3,1}=H_{1,3}$, we finally get $H_{1,2}=H_{1,3}$. Remember, $\hat H$ is unnecessary to be a hermitian operator. 
In some cases, such as $C_5$ symmetry, its traces of irreducible representations can be complex numbers and this kind of result is impossible to be generated through a real and hermitian operator and 
usually the complex conjugate terms will mix together to be a reducible representation.
The reason why I choose a real and hermitian operator for $C_{4v}$ symmetry calculation is that I want to get the representations with real numbers.
\subsubsection{get irreducible representation from eigenfunctions of $\hat{H}$}
This part is simple, you just need to find out the eigenfunctions with
the same eigenvalues which are related to the same irreducible representation
$\hat\Gamma^\alpha$ and it obeys 
\begin{align}
\hat\Gamma^\alpha(\hat R_i)\hat\Gamma^\alpha(\hat R_j)=\hat\Gamma^\alpha(\hat R_i \hat R_j). \label{eq:condition_of_rep}
\end{align}
If we set 
\begin{align*}
\sum_{m'_\alpha}\Gamma^\alpha_{m_\alpha m'_\alpha}(\hat{R})\psi_{m'_\alpha}({\bf r})&= \psi_{m_\alpha}(\hat{R}{\bf r}),\\
\psi_{m_\alpha}({\bf r})&=\sum_{\bf G}C_{m_\alpha,{\bf G}}e^{i{\bf G}\cdot{\bf r}},
\end{align*}
where $m_\alpha$ denotes the index of 
eigenfunction in $\alpha_{\rm th}$ irreducible representation
, we get 
\begin{align}
C_{m_\alpha,\hat{R}{\bf G}}&=\sum_{m_\alpha m'_\alpha}\Gamma^\alpha_{m_\alpha m'_\alpha}(\hat{R})C_{m'_\alpha,{\bf G}}, \label{eq:irr_C}
\end{align}
because of the independence of plane waves.
Since $C_{m_\alpha,{\bf G}}$ is the eigenfunction, $C_{m_\alpha,\hat{R}{\bf G}}$ is just the 
rearrangement of $C_{m_\alpha,{\bf G}}$ based on group-multiplication table.
For example, if we want to get $\hat\Gamma^\alpha(\hat{R}_3)$ of square lattice case, we need to provide Eq.~\ref{eq:irr_C} $\{C_{m_\alpha,{\bf G}}|m_\alpha,{\bf G}\}$ and $\{C_{m_\alpha,\hat{R}_3{\bf G}}|m_\alpha,{\bf G}\}$, where 
$\hat{R}_3\{{\bf G}\}=\hat{R}_3\hat{R}_{1\cdots 8}{\bf G}_0=
\left[ {\bf G}_3,{\bf G}_1,{\bf G}_4,{\bf G}_2,{\bf G}_7,{\bf G}_5,{\bf G}_8,{\bf G}_6\right]$ which is derived from the third row of group-multiplication table.
Since we get $\{C_{m_\alpha,{\bf G}}|m_\alpha,{\bf G}\}$ as the eigenfunctions with the same eigenvalues and $\{C_{m_\alpha,\hat{R}_3{\bf G}}|m_\alpha,{\bf G}\}$ is just the rearrangement of $\{C_{m_\alpha,{\bf G}}|m_\alpha,{\bf G}\}$, it is no doubt that we can get $\hat\Gamma^\alpha(\hat R_3)$.
Hence, once we get the eigenfunctions, we can get the irreducible representation from Eq.~\ref{eq:irr_C}.
\subsubsection{character table}
Since we can get the irreducible representation, what we need to do is taking 
the trace of irreducible representation is enough.

\subsubsection{one interesting behavior}
When we derive an irreducible representation by this way, we get each
irreducible representation $l_\alpha$ times where each $\hat\Gamma^\alpha(\hat R)$ is a $l_\alpha\times l_\alpha$ matrix; besides, we need $l_\alpha$ independent eigenstates with the same eigenvalue to derive $\hat\Gamma^\alpha$. 
Hence, there are $l_\alpha^2$ eigenstates are related to an irreducible representation $\hat\Gamma^\alpha$.
When we consider all the irreducible representations, we get
\begin{align}
n_R=\sum_\alpha l_\alpha^2,
\end{align}
where $n_R$ is the number of symmetry operators and $\hat H$ is a $n_R\times n_R$ matrix.
Hence, we get the relationship predicted by group theory.
The reason is shown as follows:
When we substitute Eqs.~\ref{eq:H_R&H} and ~\ref{eq:irr_C} into $\hat H\psi=\varepsilon\psi$ where $\hat H$ is the one I used to generate irreducible representations, we will get
\begin{align}
  \sum_{m'_\alpha}\sum_{\hat R} H_{{\bf G}_0,\hat R{\bf G}_0} \Gamma^\alpha(\hat R)_{m_\alpha,m'_\alpha}C_{m'_\alpha,{\bf G}_0}=C_{m_\alpha,{\bf G}_0},
\end{align}
where $\sum_{\hat R}H_{{\bf G}_0,\hat R{\bf G}_0} \hat\Gamma^\alpha(\hat R)$ will be a $l_\alpha\times l_\alpha$ matrix and its eigenvalues will be $l_\alpha$ different ones and $C_{{\bf G}_0}$ will be a $l_\alpha \times 1$ vector for each eigenvalue. Hence, we can generate $l_\alpha$ orthogonal eigenvectors for each eigenvalue through Eq.~\ref{eq:irr_C} since we know $C_{{\bf G}_0}$. This is the reason that we will get $l_\alpha^2$ states related to one irreducible representation.
In fact, we can use this way to reduce the computational time based on this concept. For example, the original matrix size of the $\hat H$ of $C_{4v}$ is $8\times 8$, now you can reduce it into four $1\times 1$ and one $2\times 2$ matrices. The detail of how to reduce the computational time for real case is not shown here.
\section{The reason why we can get all the irreducible representations by this way\label{sec:irr_prove_get_IR}}
If we want to prove this thing, we need to prove four things first.
\begin{enumerate}
\item The functions belong to different irreducible representation will orthogonal to each other.
\item If the functions of $\{\phi_i(\hat{R}{\bf r})|\hat{R}\}$ are independent to each other, all the irreducible representations can be derived from this set.
Hence, we don't need to deal with infinite number of functions to find the irreducible representations and the number of set is the same as of symmetry operators.
\item If a Hamiltonian $\hat H$ obeys $\hat H({\bf r})=\hat H(\hat{R}{\bf r})$
, the eigenfunctions with the same eigenvalue belong to the same representation of $\{\hat{R}\}$.
Hence, we can get the representations through solving eigenvalue problem.
If the unnecessary degeneracy are broken, we even can get the irreducible representations directly.
\item One kind of trace value of irreducible representation is just related to one kind of irreducible representation.
Hence, once we find the irreducible representations through one Hamiltonian, they are what we need.
\end{enumerate}

\subsection{Proof of the first part}

Assume a function set obeys
\begin{align}
\psi_{n_\alpha}(\hat{R}{\bf r})&=\sum_{n'_\alpha}\Gamma^\alpha_{n_\alpha,n'_\alpha}(\hat{R}) \psi_{n'_\alpha}({\bf r}), \label{eq:irr_expand_psi} 
\end{align}
\begin{align*}
\int d{\bf r}\psi^*_{n_\alpha}({\bf r})\psi_{n_\alpha}({\bf r})&=1, 
\end{align*}
and we want to find the value of $\int d{\bf r}\psi^*_{n_\alpha}({\bf r})\psi_{m_\beta}({\bf r})$, where $\alpha$ and $\beta$ denote the index of irreducible representations and $m$ and $n$ are the index of eigenfunctions in each irreducible representation, respectively.
Because the integration is over the whole space, we get
\[
\int d{\bf r}\psi^*_{n_\alpha}({\bf r})\psi_{m_\beta}({\bf r})=
\int d{\bf r}\psi^*_{n_\alpha}(\hat{R}{\bf r})\psi_{m_\beta}(\hat{R}{\bf r}).
\]
Now, let's expand $\psi_{n_\alpha}(\hat{R}{\bf r})$ by Eq.~\ref{eq:irr_expand_psi} and consider
\begin{align}
\sum_{\hat{R}}\Gamma^{\alpha*}_{n1_\alpha,n1'_\alpha}(\hat{R})\Gamma^\beta_{m2_\beta,m2'_\beta}(\hat{R})
=\frac{n_R}{l_\alpha}\delta_{\alpha,\beta}\delta_{n1_\alpha,m2_\beta}\delta_{n1'_\alpha,m2'_\beta}, \label{eq:irr_orthogonal_D}
\end{align}
which is shown in the textbook,
we will get
\begin{align}
&\int d{\bf r}\psi^*_{n1_\alpha}({\bf r})\psi_{m2_\beta}({\bf r})=
\frac{1}{n_R}\sum_{\hat{R}}\int d{\bf r}\psi^*_{n1_\alpha}(\hat{R}{\bf r})\psi_{m2_\beta}(\hat{R}{\bf r}) \nonumber\\
&=
 \sum_{\hat{R}}\Gamma^{\alpha*}_{n1_\alpha,n1'_\alpha}(\hat{R})\Gamma^\beta_{m2_\beta,m2'_\beta}(\hat{R})
 \sum_{n1'_\alpha,m2'_\beta}  \frac{ \int d{\bf r}\psi^*_{n1'_\alpha}({\bf r})\psi_{m2'_\beta}({\bf r}) }{n_R}
 \nonumber\\
&=\delta_{\alpha,\beta}\delta_{n1_\alpha,m2_\beta},
\end{align}
where $n_R$ and $l_\alpha$ are the number of symmetry operators and of $n_\alpha$ which means $n_\alpha=1\to l_\alpha$, respectively.
Hence, they are orthogonal to each other.

\subsection{ Proof of the second part}

Assume all the functions can be expanded by $\{\phi_m({\bf r})|m\}$,
where $\{\phi_m({\bf r})|m\}$ are independent to each other and 
for arbitrary $n$ and $\hat{R}$, $\phi_n(\hat{R}{\bf r})\subset\{\phi_m({\bf r})|m\}$. 
For example, plane waves obey this requirement for rotational symmetry in periodic structure.
Hence, we can define that $\phi^\Delta_i({\bf r})$ obeys
\begin{align*}
\{\phi^\Delta_i({\bf r})|i\} &\subset \{\phi_m({\bf r})|m\} \\
\{\phi^\Delta_i(\hat{R}{\bf r})|i,\hat{R}\}&= \{\phi_m({\bf r})|m\}, 
\end{align*}
and $\{\phi^\Delta_i({\bf r})|i\}$ is a set with minimal number of members.
Therefore, an arbitrary function $\psi_{n_\alpha}({\bf r})$ can be written as 
\begin{align}
\psi_{n_\alpha}({\bf r})=\sum_{i,\hat{R}}C^i_{n_\alpha}(\hat{R})\phi^\Delta_i(\hat{R}^{-1}{\bf r}), \label{eq:expand_psi}
\end{align}
where $n_\alpha$ means the $n_{\rm th}$ states of $\alpha_{\rm th}$ irreducible representation.
If we let 
\begin{align*}
\psi_{n_\alpha}(\hat{R}_1{\bf r})=\sum_{n'_\alpha}\Gamma^\alpha_{n_\alpha,n'_\alpha}(\hat{R}_1) \psi_{n'_\alpha}({\bf r}),
\end{align*}
expand $\psi$ by Eq.~\ref{eq:expand_psi} and use the concept of $\hat{R}\{\hat{R}\}=\{\hat{R}\}$, we have
\begin{align*}
\sum_{i,\hat{R}}C^i_{n_\alpha}(\hat{R}_1\hat{R})\phi^\Delta_i(\hat{R}^{-1}{\bf r})=\sum_{n'_\alpha,i,\hat{R}}\Gamma^\alpha_{n_\alpha,n'_\alpha}(\hat{R}) C^i_{n'_\alpha}(\hat{R})\phi^\Delta_i(\hat{R}^{-1}{\bf r}).
\end{align*}
Because of the independence of $\{\phi^\Delta_i({\bf r})|i\}$,
we can get 
\begin{align}
\sum_{\hat{R}}C^i_{n_\alpha}(\hat{R}_1\hat{R})\phi^\Delta_i(\hat{R}^{-1}{\bf r})=\sum_{n'_\alpha,\hat{R}}\Gamma^\alpha_{n'_\alpha,n_\alpha}(\hat{R}_1) C^i_{n'_\alpha}(\hat{R})\phi^\Delta_i(\hat{R}^{-1}{\bf r}), \label{eq:irr_finding_D}
\end{align}
which means the representations can be just related to $i_{\rm th}$ subset
$\{\phi^\Delta_i(\hat{R}{\bf r})|\hat{R}\}$ .
For finding all the irreducible representations, we need to require that the elements of $\{\phi^\Delta_i(\hat{R}{\bf r})|\hat{R}\}$ are independent to each other.
If not, it is possible $C^i_{n_\alpha}(\hat{R})=0$ for all $\hat{R}$.
Besides, when we require that $\{\phi^\Delta_i(\hat{R}{\bf r})|\hat{R}\}$ are independent to each other, Eq.~\ref{eq:irr_finding_D} can be simplified to be
\begin{align}
C^i_{n_\alpha}(\hat{R}_1\hat{R})=\sum_{n'_\alpha}\Gamma^\alpha_{n_\alpha,n'_\alpha}(\hat{R}_1) C^i_{n'_\alpha}(\hat{R}), \label{eq:irr_finding_D2}
\end{align}
that's why we get Eq.~\ref{eq:irr_C}. Here we can find that the formula of bases is not important. 

Besides, in the following, when we provide $C^i_{n_\alpha}$ for Eq.~\ref{eq:irr_finding_D2}, we can prove that we just can find one kind of $\Gamma^\alpha$ when they are independent to each other and $l_\alpha\le n_R$, where each $\hat\Gamma^\alpha(\hat R)$ is a $l_\alpha\times l_\alpha$ matrix.
Please note that $C^i_{n_\alpha}(\hat{R})$ are independent to each other for different $n_\alpha$ because of the first proof.
If they are not independent to each other, we can prove that the related $\hat\Gamma^\alpha$ derived from Eq.~\ref{eq:irr_C} is reducible. The proof is as follows:

If we define $\left|C^i_{n_\alpha}\right>$ and $\left|C^i_{n_\alpha}(\hat{R}_1)\right>$ as $C^i_{n_\alpha}(\hat{R})$ and 
$C^i_{n_\alpha}(\hat{R}_1\hat{R})$ and they are $n_R\times 1$ vectors,
Eq.~\ref{eq:irr_C} can be written as
\begin{align*}
\left( \left|C^i_1\right> \left|C^i_2\right> \cdots \right) \hat\Gamma^T(\hat{R}_1) =
\left( \left|C^i_1(\hat{R}_1)\right> \left|C^i_2(\hat{R}_1)\right> \cdots \right), 
\end{align*}
and 
\begin{align}
\left( \left|C^i_1\right> \left|C^i_2\right> \cdots \right) & \hat\chi \hat\chi^{-1}\hat\Gamma^T(\hat{R}_1) \hat\chi = \nonumber\\ 
& \left( \left|C^i_1(\hat{R}_1)\right> \left|C^i_2(\hat{R}_1)\right> \cdots \right) \hat\chi, \label{eq:irr_C_with_chi}
\end{align}
where $\Gamma^T_{i,j}=\Gamma_{j,i}$ and $\hat\chi$ is an arbitrary $l_\alpha\times l_\alpha$ matrix whose determine is not zero.
If $\{\left|C^i_{n_\alpha}\right>|n\}$ is a dependent set, we always can find a $\hat\chi$ whose determine is not zero and 
\[
\left( \left|C^i_1\right> \left|C^i_2\right> \cdots \right) \hat\chi =
\left( \left|C'^i_1\right> \cdots \O \right).
\]
Besides, if $\left|C'^i_{l_\alpha}\right>=\O$, because of rearrangement, $|C'^i_{l_\alpha}(\hat{R})>=\O$.
Equation~\ref{eq:irr_C_with_chi} will become
\[
\left( \left|C'^i_1\right> \cdots \O \right)\hat\Gamma'^T(\hat{R}_1)=
\left( \left|C'^i_1(\hat{R}_1)\right> \cdots \O \right),
\]
where $\hat\Gamma'^T(\hat{R}_1)=\hat\chi^{-1}\hat\Gamma^T(\hat{R}_1)\hat\chi$.
Because of this equation, when we choose $\hat\Gamma(\hat{R}^{-1})=\hat\Gamma(\hat{R})^\dagger$ and $\hat\chi^{-1}=\hat\chi^\dagger$,
we get $\Gamma'_{m,l_\alpha}=\Gamma'_{l_\alpha,m}=0$ where $m$ is $1$ to $l_\alpha-1$
.
That means $\hat\Gamma$ is reducible.
Hence, if $l_\alpha$ is larger than the number of 
symmetry operators, $\{\left|C^i_{n_\alpha}\right>|n\}$ must be a dependent set.

If the elements of $\{\left|C^i_{n_\alpha}\right>|n\}$ are independent to each other, this set will
just be related to a unique representation.
The simple proof is as follows:
If  $\hat\Gamma$ and $\hat\Gamma'$ obey Eq.~\ref{eq:irr_finding_D2},
we get
\begin{align*}
\O &=\left( \left|C^i_1\right> \cdots \right) (\hat\Gamma^T-\hat\Gamma'^T).
\end{align*}
Because the elements of $\{\left|C^i_{n_\alpha}\right>|n\}$ are independent to each other, $\hat\Gamma-\hat\Gamma'=\O$.
Therefore, we just can get one kind of $\hat\Gamma$ when $\left|C^i_{n_\alpha}\right>$ are provided.

Hence, we can find the irreducible representations from $\{\phi^\Delta_i(\hat{R}{\bf r})|\hat{R}\}$ whose number equals of symmetry operators.

Here, let me summarize this proof. 
Assume a function set used to find the representation are described by $\{C^i_{n_\alpha}(\hat{R})|i,n,\hat{R}\}$.
Based on the first proof, $\{C^i_{n_\alpha}(\hat{R})|i,n,\hat{R}\}\equiv\{\left|C_{n_\alpha}\right>|n\}$ must be independent to each other where
$\left|C_{n_\alpha}\right>$ is a $n_R n_i\times 1$ vector, $n_R$ and $n_i$ denote the number of symmetry operators and of $\phi^\Delta_i$, respectively. 
If the elements of $\{C^0_{n_\alpha}(\hat{R})|n,\hat{R}\}$ are already independent to each other, we can get an unique $\hat\Gamma$ from this set and ignore all other contributions from $i\neq 0$ terms.
That's why we just need $\{\phi^\Delta_i(\hat{R}{\bf r})|\hat{R}\}$ to find the irreducible representations.


\subsection{ Proof of third part}

The problem we want to study is 
\begin{align}
\hat H({\bf r})\psi_{n_\alpha}({\bf r})=\varepsilon_\alpha\psi_{n_\alpha}({\bf r}),
\end{align}
where $\hat H({\bf r})$ is provided, and $\varepsilon_\alpha$ and $\psi_{n_\alpha}$
are the eigenvalue and eigenstate of $\hat H({\bf r})$.
Because $\hat H({\bf r})=\hat H(\hat{R}{\bf r})$, we have 
\[
\hat H({\bf r})\psi_{n_\alpha}(\hat{R}{\bf r})=\varepsilon_\alpha\psi_{n_\alpha}(\hat{R}{\bf r}),
\]
and that means $\{\psi_{n_\alpha}(\hat{R}{\bf r})|\hat{R}\}$ are degenerate states.
Hence, $\psi_{n_\alpha}({\hat R}{\bf r})=\sum_{n'_\alpha}\Gamma^\alpha_{n_\alpha,n'_\alpha}(\hat{R}) \psi_{n'_\alpha}({\bf r})$ , and $\hat\Gamma^\alpha$ will be one kind of representation because it obeys Eq.~\ref{eq:condition_of_rep}.
Therefore, for the states with the same eigenvalue $\varepsilon_\alpha$, we should get the representation from them. If the unnecessary degenerate are broken, 
they are related to the irreducible representation.


\subsection{ Proof of the fourth part}

It is easy to prove that the same irreducible representations will relate to the same trace.
The same means that they are $\{\hat\chi^{-1}\hat\Gamma \hat\chi|\det\left(\hat\chi\right)\neq 0\}$, and their trace will be the same because of ${\rm Tr}\left(\hat\chi^{-1}\hat\Gamma \hat\chi\right)={\rm Tr}\left(\hat\Gamma\right)$.
If we want to prove that the same trace will be correspond to the same irreducible representations, we can prove it through the orthogonality of trace
, i.e. 
\begin{align}
\sum_R \hat\chi^{\alpha*}(\hat{R}) \hat\chi^\beta(\hat{R})=n_R\delta_{\alpha,\beta},
\label{eq:irr_orthogonal_chi}
\end{align}
where $\hat\chi^\beta(\hat{R})=\sum_j \Gamma^\beta_{j,j}(\hat{R})$ is the trace of $\beta_{\rm th}$ irreducible representation.
It is easy to prove this equation because the left hand side of Eq.~\ref{eq:irr_orthogonal_chi} can be expanded as 
\begin{align*}
\sum_R\hat\chi^{\alpha*}(\hat{R})\hat\chi^\beta(\hat{R}) &=
\sum_{i,j,\hat{R}}\Gamma^{\alpha*}_{i,i}(\hat{R})\Gamma^\beta_{j,j}(\hat{R}),
\end{align*}
and it is the same as
$ 
 \sum_{i,j}\delta_{i,j}\delta_{\alpha,\beta}\frac{n_R}{l_\alpha}=n_R\delta_{\alpha,\beta},
$ and
finally we get Eq.~\ref{eq:irr_orthogonal_chi}.

Therefore, the first and second proof tell us that we can get irreducible representations through a set of independent functions; the third proof tell us that we can get this independent functions as the eigenfunctions of a hamiltionian which obeys the symmetry and the fourth part help us to classify the irreducible representations.

\section{Conclusion~\label{sec:conclusion}}
Although the proof is based on the real space, the information we need is just the group-multiplication table as shown in Sec.~\ref{sec:irr_get_IR}. 
Hence, once we know the group-multiplication table, the related irreducible representations can be obtained and it is independent of what kind of bases we choose.
Since that, it is not restricted in Lie group. 
The code of this paper is submitted to matlab and welcome you to test it. In that case, I use Buckminsterfullenerene (C60) whose symmetry is icosahedral-inversion symmetry ~\cite{icosahedral_B_R_JUDD} as an example and get its character table as
\begin{widetext}
\begin{align*}
\begin{array}{rrrrrrrrrr}
1E&	15{C_2'}&	12{C_5^2}&	12{C_5^4}&	15{C_2}&	12{{C_5^4}'}&	20{C_3^2}&	20{{C_3^2}'}&	12{{C_5^2}'}&	1i\\
\hline\\
1&	-1&	1&	1&	1&	-1&	1&	-1&	-1&	-1\\
3&	-1&	1.618&  -0.618&  -1&	-0.618&	0&	0&	1.618&	3\\
5&	-1&	0&	0&	1&	0&	-1&	1&	0&	-5\\
3&	-1&	-0.618&	1.618&	-1&	1.618&	0&	0&	-0.618&	3\\
3&	1&	1.618&	-0.618&	-1&	0.618&	0&	0&	-1.618&	-3\\
5&	1&	0&	0&	1&	0&	-1&	-1&	0&	5\\
4&	0&	-1&	-1&	0&	1&	1&	-1&	1&	-4\\
4&	0&	-1&	-1&	0&	-1&	1&	1&	-1&	4\\
3&	1&	-0.618&	1.618&	-1&	-1.618&	0&	0&	0.618&	-3\\
1&	1&	1&	1&	1&	1&	1&	1&	1&	1
\end{array}
\end{align*}
\end{widetext}
where the first row denotes the number of elements of each class with the class type and the following is the trace of each irreducible representation. Besides, the values shown as $-0.618$ and $0.618$ are $(1-\sqrt{5})/2$ and $(1+\sqrt{5})/2$ with error about $10^{-16}$, respectively.
\acknowledgments
I would like to acknowledge NCTS and the financial support from NSC of Taiwan under Grant No. NSC 95-2745-M-006-004-.
\bibliography{my_study}
\end{document}